# "Zero change" platform for monolithic back-end-of-line integration of phase change materials in silicon photonics

**Maoliang Wei[1,†], Kai Xu[1,†], Bo Tang[2,†], Junying Li[1,3,*], Yiting Yun[1], Peng Zhang[2], Yingchun Wu[4,5] Kangjian Bao[4,5], Kunhao Lei[1], Zequn Chen[4,5], Hui Ma[1], Chunlei Sun[4,5], Ruonan Liu[2], Ming Li[6,*], Lan Li[4,5,*], Hongtao Lin[1,*]**

[1] State Key Laboratory of Modern Optical Instrumentation, Key Laboratory of Micro-Nano Electronics and Smart System of Zhejiang Province, College of Information Science and Electronic Engineering, Zhejiang University, Hangzhou 310027, China

[2] Institute of Microelectronics of the Chinese Academy of Sciences, Beijing 100029, China

[3] Hangzhou Institute for Advanced Study, University of Chinese Academy of Sciences, Hangzhou 310024, China

[4] Key Laboratory of 3D Micro/Nano Fabrication and Characterization of Zhejiang Province, School of Engineering, Westlake University, Hangzhou, Zhejiang 310030, China

[5] Institute of Advanced Technology, Westlake Institute for Advanced Study, Hangzhou, Zhejiang 310024, China

[6] State Key Laboratory on Integrated Optoelectronics, Institute of Semiconductors, Chinese Academy of Sciences, Beijing 100083, China

[†] These authors contributed equally to this work.

* Corresponding author: hometown@zju.edu.cn; junyingli@ucas.edu.cn; lilan@westlake.edu.cn; ml@semi.ac.cn

**Abstract**

Monolithic integration of novel materials for unprecedented device functions without modifying the existing photonic component library is the key to advancing heterogeneous silicon photonic integrated circuits. To achieve this, the introduction of a silicon nitride etching stop layer at selective area, coupled with low-loss oxide trench to waveguide surface, enables the incorporation of various functional materials without disrupting the reliability of foundry-verified devices. As an illustration, two distinct chalcogenide phase change materials (PCM) with remarkable nonvolatile modulation capabilities, namely $Sb_2Se_3$ and $Ge_2Sb_2Se_4Te_1$, were monolithic back-end-of-line integrated into silicon photonics. The PCM enables compact phase and intensity tuning units with zero-static power consumption. Taking advantage of these building blocks, the phase error of a push-pull Mach-Zehnder interferometer optical switch could be trimmed by a nonvolatile phase shifter with a 48% peak power consumption reduction. Mirco-ring filters with a rejection ratio >25dB could be applied for >5-bit wavelength selective intensity modulation, and waveguide-based >7-bit intensity-modulation photonic attenuators could achieve >39dB broadband attenuation. The advanced "Zero change" back-end-of-line integration platform could not only facilitate the integration of PCMs for integrated reconfigurable photonics but also open up the possibilities for integrating other excellent optoelectronic materials in the future silicon photonic process design kits.

**Introduction**

Silicon photonics holds broad prospects for practical applications in high-speed optical communication[1], microwave photonics[2], optical neural networks[3], and optical quantum computing[4]. Accurate modulation of waveguide refractive index and absorption is critical for effectively implementing functional units in silicon photonics, facilitating precise control and adjustment of functionality within silicon photonic networks[5]. However, current modulation schemes in photonic chips primarily rely on modulation approaches such as thermo-optic modulation and free-carrier dispersion in silicon, exhibiting weak modulation strength (usually with the change of effective refractive indices $\Delta n_{eff} < 10^{-3}$) and necessitating a continuous power supply[6]. This leads to large device sizes and high static power consumption in photonic chips, limiting progress in large-scale optoelectronic integration.

Introducing innovative materials for function units in silicon photonics has become imperative to attain exceptional device performance and reduce power consumption[7]. Various materials such as electro-optic polymer[8], metal-insulator-transition oxide[9], and 2D materials[10] had been integrated for ultra-compact or ultra-fast volatile light modulation. An ongoing trend to integrating nonvolatile modulated materials such as charge-trapping materials[11,12], ferroelectric materials[13], and chalcogenide phase change materials[14] is crucial for lowering the static power consumption of reconfigurable photonic circuits[15]. Despite the significant progress in prototype devices with exceptional performance, the fabrication process flows are incompatible with the existing silicon photonics foundry process, rendering the established passive and active photonic component design kits unsuitable for direct application[16].

Enabling the monolithic integration of functional materials into silicon photonics while maintaining the existing fabrication process and using the available process design kits (PDKs) is of utmost importance[15]. Chalcogenide phase change materials, such as an example, can be directly deposited on silicon and have attracted significant attention thanks to their nonvolatile properties[17-20], making them promising candidates for compact (~10 μm) and zero-static power photonic devices. In the past decades, a plethora of PCM-integrated reconfigurable photonic devices have been extensively developed for intensity modulation[21-29], phase tuning[30-33], and light path switching[34,35]. Moreover, they play a crucial role in constructing photonic networks and serve as essential elements for optical storage[36], in-memory computing[37], and analog optical computing[38,39]. Despite the significant advancements in PCM-integrated photonics, the full compatibility of integrating PCMs into the entire silicon photonics fabrication flow remains highly challenging. Therefore, an imperative back-end integration approach is urgently needed to facilitate the post-processing integration of PCMs, thereby enhancing the feasibility of large-scale integration of nonvolatile reconfigurable optoelectronic chips.

In this paper, a "Zero change" platform for a monolithic back-end-of-line integration scheme facilitating the large-scale integration of PCM-based photonic devices was demonstrated. By customizing the full process flow of silicon photonics and introducing a CMOS-compatible silicon nitride (SiN) layer as an etch stop layer on Si waveguides, a deep $SiO_2$ trench was etched with a low insertion loss of < 0.09 dB/trench for subsequent integration of various PCMs. Two kinds of chalcogenide phase change materials (PCMs), $Sb_2Se_3$ and $Ge_2Sb_2Se_4Te_1$(GSS4T1), with completely different nonvolatile modulation capabilities, were monolithically back-end-of-line integrated into silicon photonics. Electrical-assisted programmability of the fabricated devices was verified, enabling reconfigurable post-trimming, multi-level nonvolatile phase modulation, and intensity modulation in optoelectronic chips. This endeavor not only showcased the back-end-of-line integration technique for combining phase change memory (PCM) with silicon photonics but also accomplished this without the need to modify the existing library of passive and active photonic components. Furthermore, it establishes a clear path for integrating other promising optoelectronic materials into future silicon optoelectronic chips.

**Results**
**Back-end-of-line integration of PCMs into commercial silicon photonics platform**
Large-scale fabrication based on the CMOS platform without modifying the existing passive and active photonic component library is essential for realizing various practical applications of PCM-based nonvolatile electrically programmable photonic chips. However, PCM is a kind of material that is incompatible with standard CMOS processes. Therefore, a back-end-of-line integration of PCMs is coveted for commercial photonics platform-compatible nonvolatile devices. Here, we propose a trench etch process utilizing SiN as the etch stop layer to realize deep $SiO_2$ cladding etching above the functional areas (where PCM would be deposited), which is suitable for developing CMOS back-end integration of multiple functional materials[40].

The SiN-assisted silicon photonic process was conducted on a 200-mm wafer at the IMECAS foundry, including the low-loss $SiO_2$ trench etching atop the functional areas. The integration of PCM was accomplished by a back-end-of-line process involving ultraviolet lithography and film deposition, which exhibits significant potential for facilitating large-scale integration. The customized silicon photonic process incorporating a layer of SiN serves as an etch stop layer to preserve the Si waveguide from damage during the etching of $SiO_2$ trenches (see Fig. 1(a)). Firstly, the patterning of photonic devices, implantation, and ion activation was implemented on a 200-mm SOI wafer comprising a 220-nm silicon layer on top of a 2-μm buried oxide layer. Secondly, a sequential deposition of a 5-nm $SiO_2$ and a 20-nm SiN was performed, followed by lithography and etching techniques to fabricate the etch stop layer. Following the deposition of $SiO_2$ and subsequent metal interconnection, the silicon oxide above the waveguides in functional areas of the photonic devices was selectively etched. Due to the high etching selection ratio (>50:1) between $SiO_2$ and SiN, the etching process is effectively halted at the etch stop layer (SiN), thereby preventing damage to the silicon waveguide. Finally, a wet etch process was employed to remove the SiN layer above the functional areas.

Benefiting from the etch stop layer, waveguides with low-loss narrow $SiO_2$ trenches were realized, with their measured transmission spectra shown in Fig. 1(b), where the numbers of cascaded trenches are 10, 20, and 30, respectively. The cut-back measurement suggested that the insertion loss introduced by a single trench is as low as 0.083 dB at 1550 nm, as shown in Fig. 1(c). The low-loss trenches provide a convenient approach for the back-end integration of PCMs, facilitating compatibility with available silicon photonic devices in commercially available PDKs. To prevent any performance degradation of the device, the post-processing temperature should be lower than 450 ℃[41]. Additionally, post-fabrication patterning was accomplished exclusively by ultraviolet lithography, thereby possessing the potential for large-scale integration with high throughput. The detailed fabrication process flowchart for the back-end integration of PCMs is illustrated in Fig. 1(a). The PCM thin film was deposited through magnetron sputtering followed by a lift-off process upon the $SiO_2$ cladding trench window (process temperature <150 ℃). Afterward, the chip was annealed at 200 ℃ and 300 ℃ in a nitrogen atmosphere for 15 minutes to facilitate the crystallization of $Sb_2Se_3$ and $GeSbSe_4Te_1$, respectively. A 30 nm-thick $Al_2O_3$ protective layer was then deposited by atomic layer deposition (process temperature < 150℃). Finally, the contact window above the metal electrode was etched to ensure electrical interconnection (process temperature < 110℃). Fig. 1(d) illustrates the functional region's structure after fabrication. The electro-thermal control of the PCM induced by the PIN diode enables a nonvolatile response of the effective refractive index in the hybrid waveguides.

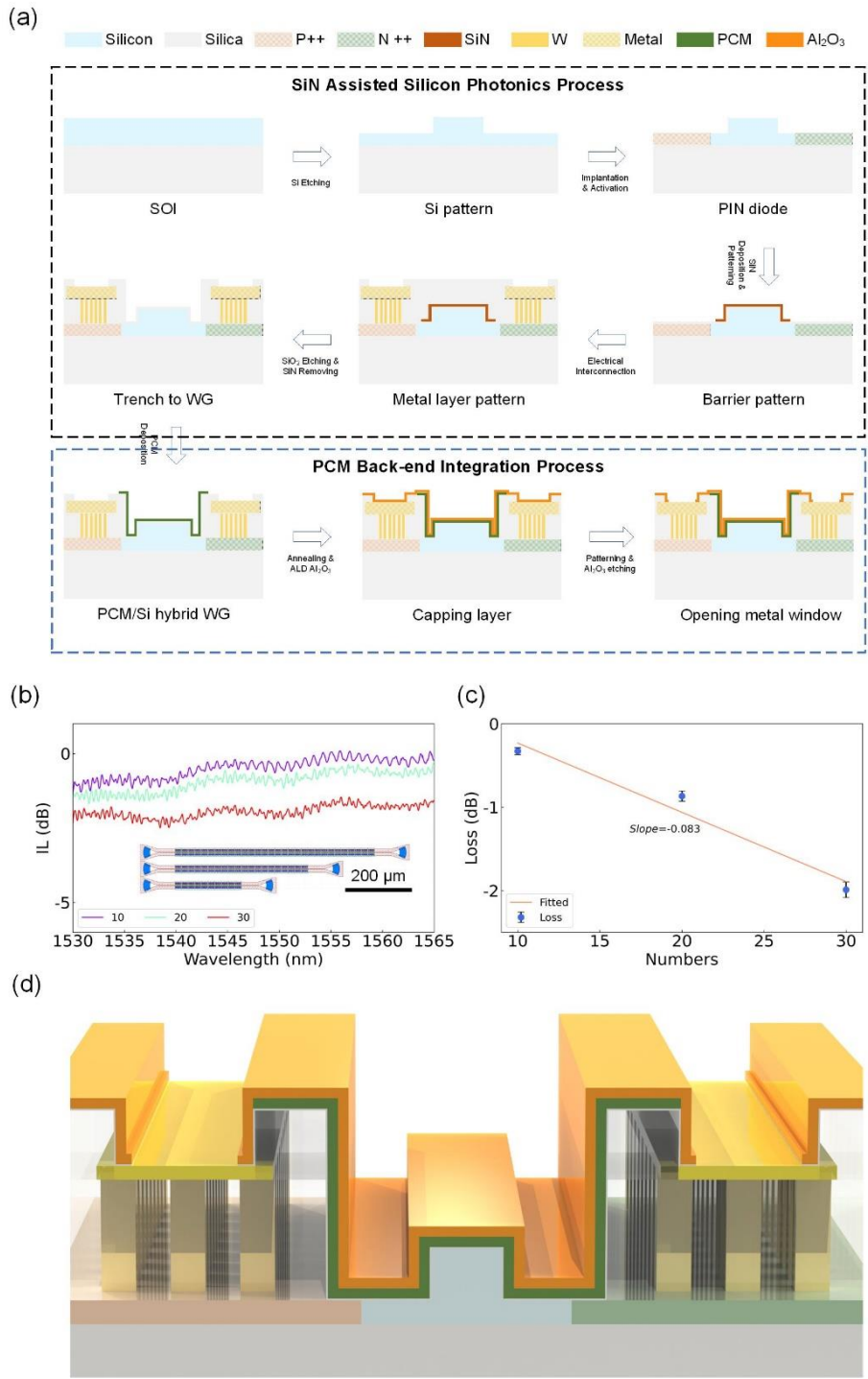

Fig. 1 SiN-assisted silicon photonic process for the back-end-of-line integration of PCMs. (a) Flowchart of the device fabrication process. (b) Measured transmission spectra of waveguides with different numbers of cascaded trenches. The inset is the layout of the cascaded devices. (c) Assessed waveguide loss introduced by $SiO_2$ trench etch process employing cut-back method. (d) 3D schematic image of the device after fabrication.

**Back-end integrated SbSe enabled trimming of push-pull MZI switch**

Silicon photonic devices inevitably suffer from fabrication errors and consequent deviation of performance from the intended design. Meanwhile, the device-to-device deviations could be significantly magnified in networks, leading to an increased complexity when configuring PICs. Post-fabrication trimming (PFT) enables the calibration of the photonic devices after their fabrication. Compared to other trimming methods, including femtosecond-laser annealing[42] and Ge ion implant-and-annealing[43,44], active and reversible trimming enabled by PCM manipulation[30] possesses a significantly improved degree of freedom. In this section, we demonstrate the post-fabrication active trimming technique by electrically fine-tuning a low-loss $Sb_2Se_3$ patch integrated using back-end integration.

The schematic diagram in Fig. 2(a) illustrates the device structure and operational principle of low-loss PCM-based PFT, exemplified by a push-pull MZI switch. A section of p-i-n doped Si waveguide, covered with a patch of $Sb_2Se_3$, was employed as a trimming unit. An identical structure was set on another branch of the MZI to achieve optical loss equilibrium. During the PFT, a sequence of electrical pulses is applied to the trimming unit, thereby triggering the amorphization of $Sb_2Se_3$ and inducing a nonvolatile change of the refractive index in the $Sb_2Se_3$/Si hybrid waveguide. Therefore, the optical power between the two output ports gradually achieved equilibrium, accompanied by the balance of drive voltage.

The microscope image of the push-pull MZI is presented in Fig. 2(b). Before PFT, as shown in Fig. 2(c), the optical power-voltage (O-V) curve suggested that the optical power splitting ratio between two output ports was larger than 6 dB at 0 V, the drive voltage for the bar state(cross state) was -1.03 V (0.93 V), corresponding to a power consumption of 4.55 mW (1.00 mW). Additionally, a significant optical power imbalance spanning the C-band was observed, as shown in Fig. 2(d). Therefore, the insertion loss of the bar and cross states exhibited significant disparities due to the inherent losses associated with carrier injection-based phase modulation. (>0.4 dB at 1550 nm. The measured spectra can be found in Fig. S1(a) of SI.1).

After applying a series of electrical pulses with varying amplitudes (ranging from 3 V to 6 V with an interval of 0.01 V) and a fixed pulse duration of 500 ns, precisely control over the amorphous-and-crystalline-mixing state of $Sb_2Se_3$ was achieved, thereby enabling device trimming without impact on the electrical property of the device (see Fig. S1(b) in SI.1). The imbalance between the two ports was reduced to 0.110 dB at 0 V, as depicted by the O-V curve in Fig. 2(e). The bar and cross states' drive voltage was reduced to -0.98 V (yielding a power consumption of 2.33 mW) and 0.97 V (yielding a power consumption of 2.01 mW), respectively. This thereby reduced the voltage disparity from 0.1 V to 0.01 V, as well as improved total and peak power efficiency by >20% and >48%, respectively. Moreover, the trimming unit effectively equalized the splitting ratio of both output ports across the entire C-band (see Fig. 2(f)). The imbalance of insertion loss at the bar (2.69 dB) and cross (2.62 dB) states was effectively minimized to a mere 0.07 dB at 1550 nm while simultaneously ensuring an extinction ratio >20 dB in both states (see Fig. 2(g)). The insertion loss of the trimming unit is 0.46 dB, estimated using the cut-back method, as detailed in SI.2. The characterization of the push-pull MZI without $Sb_2Se_3$ also confirms that the primary source of insertion loss does not originate from the trimming unit(see SI. 3), but primarily from the propagation loss of the Si waveguide and the mode mismatch between the straight and bend sections. The trimmed MZI exhibits a high switching speed with 10-90% rise and 90-10% fall times of 9.26 ns and 9.81 ns, respectively (see Fig. S1(c) in SI.1). Additionally, the on-off switching ($>10^7$) and storage at room temperature (12 days) has no significant impact on the optical performance of the device, indicating that the $Sb_2Se_3$-based trimming unit is a reliable method for PFT (refer to SI.1 for detailed measurement results).

    The $Sb_2Se_3$-based trimming unit provides a reliable PFT technique, which not only achieves a balance of the drive voltage and insertion loss but also significantly reduces both the total and peak power consumption of a push-pull MZI. The embedded trimming units are essential for simplifying the control and reducing the power consumption of very large-scale PICs.

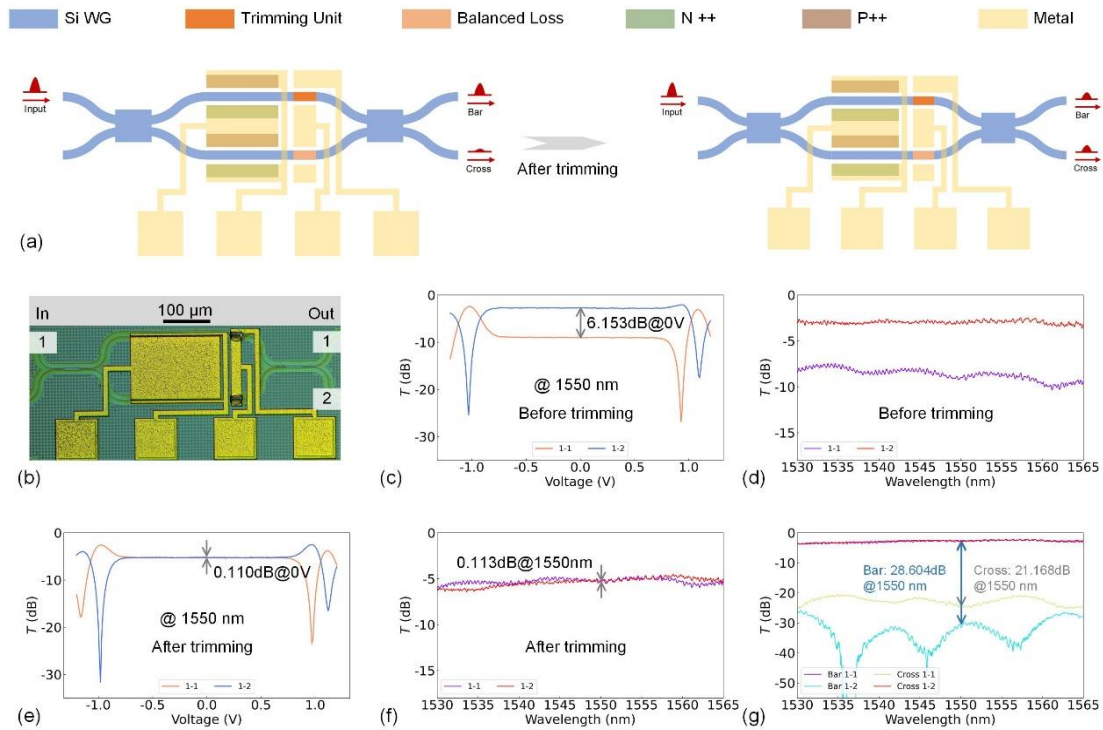

Fig. 2 Principle and performance of the push-pull MZI switch with PFT capability. (a) Principle of PFT in push-pull MZI switch. The splitting ratio of the two output ports is balanced at 0 V through precise trimming. (b) Microscope image of the device. (c) Measured optical power–voltage (O-V) curves and (d) measured spectra before PFT. (e) O-V curves and (f) measured spectra after PFT. (g) Measured spectra of bar and cross states after PFT.

**Reconfigurable nonvolatile multi-level low-loss phase modulation**

The compact nonvolatile multi-level phase modulation not only mitigates static power consumption but also enhances integration density, making it a promising candidate for constructing reconfigurable PICs such as microwave photonics[45], quantum computing[46], and coherent optical computing[47]. Here, we demonstrated a low-loss multi-level phase modulation using a back-end integrated SbSe/Si hybrid waveguide in a microring resonator(MRR).

The nonvolatile MRR switch, featuring an 8 μm-long SbSe/Si hybrid waveguide, is shown in Fig. 3(a). A ~25-nm SbSe patch was adopted to mitigate mode mismatch loss arising from the interface between the bare silicon waveguide and the SbSe/Si hybrid waveguide (see SI.4 for a detailed analysis of loss induced by mode mismatch). The reversible switching was achieved by applying a 7.15 V/500 ns pulse for amorphization and a 2.25 V/100 ms pulse for crystallization, resulting in an extinction ratio (ER) larger than 25 dB at 1551.513 nm (see Fig. 3(b)). The observed change in the measured spectra indicates that a phase shift of ~0.3π was achieved, accompanied by a crystallization-induced loss of 0.0223 dB/μm.

To achieve multi-level switching, two types of manipulating pulses were employed, including electrical pulses with varying amplitude and fixed duration (PVAFD) and pulses with varying duration and fixed amplitude (PVDFA). The change in transmittance ($\Delta T$) at 1551.145 nm was utilized for monitoring the nonvolatile multi-level phase modulation exhibited by the MRR. As shown in Fig. 3(c), 36-level (>5-bit) crystallization was realized by applying PVAFD with an interval of 0.01 V. However, achieving higher levels of nonvolatile switching using PVAFD with an amplitude interval smaller than 0.01 V poses a significant challenge. Hence, the PVDFA exhibits a potential for accommodating more distinguishable states during inducing multi-level crystallization due to the presence of numerous pulses (nearly $10^5$ at a resolution of 1 μs) with durations < 100 ms. Here, a quasi-continuous switching was achieved by applying PVDFA with gradually increased duration (see Fig. 3(d)). The demonstration of 500 switching events was also presented (see Fig.3(e)).

The back-end integrated SbSe/Si hybrid waveguide offers a reversible low-loss phase modulation, enabling fine multi-level switching by employing PVDFA, providing an attractive fine-phase tuning solution for large-scale PCM-driven photonic networks.

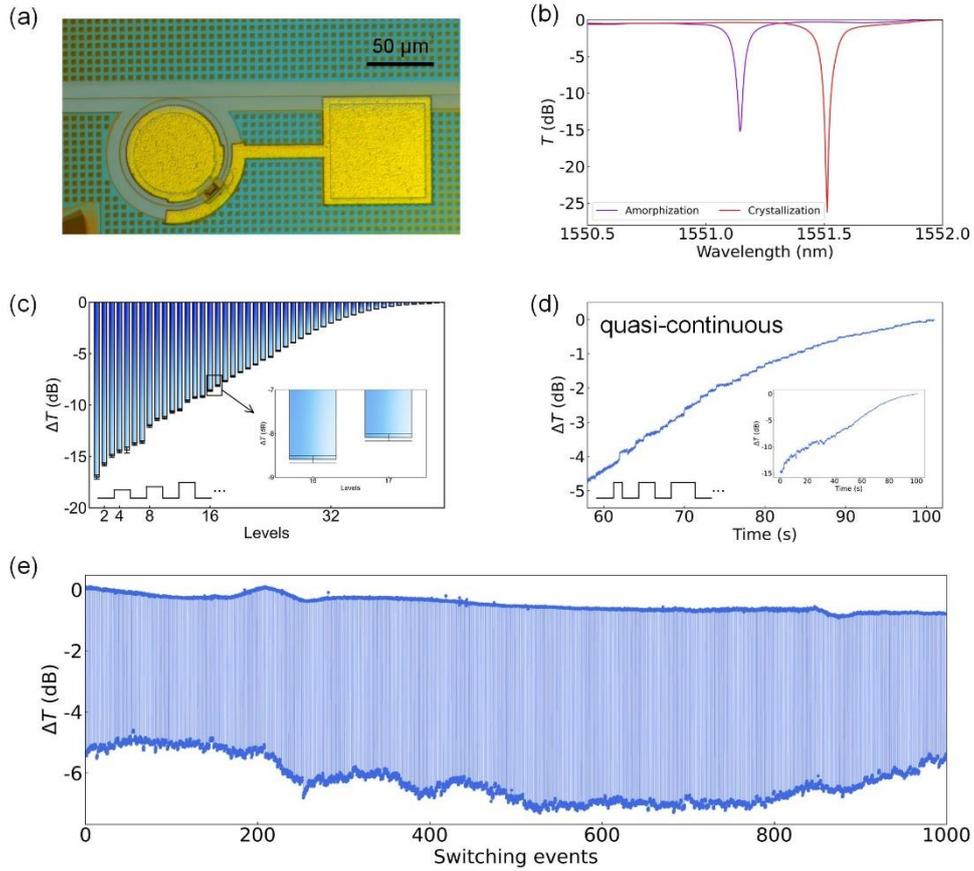

Fig. 3 The back-end integration of SbSe with MRR for multi-level phase modulation. (a) Microscope image of the MRR. (b) Measured spectra of the reversible switching events. (c) Multi-level crystallization process (ΔT at 1551.145 nm) induced by applying PVAFD. The inset shows the zoomed-in error bar of two distinct states. (d) Multi-level crystallization process (ΔT at 1551.145 nm) induced by applying PVDFA. (e) 500 reversible switching events of the device. The transmittance change (ΔT) is the difference in optical power at 1551.513 nm.

**Reconfigurable nonvolatile multi-level intensity modulation**

Intensity modulation has been widely used in optical computing[48], optical communication[49], and microwave photonics[50]. GSS4T1, a novel material possessing a low-loss amorphous state and a lossy crystalline state at telecom wavebands[19] emerges as a promising candidate for nonvolatile intensity modulation when compared with traditional GST material. Here, we pioneered the demonstration of an electrically programmable waveguide-integrated broadband optical attenuator employing back-end integrated GSS4T1.

The back-end integrated photonic attenuator offers a high extinction ratio with a small footprint (16 μm), owing to the high extinction coefficient contrast ($\Delta k$=0.549) between different states of our sputtered GSS4T1 films (see Fig. 4(a)). The microscope image of the fabricated device is shown in Fig. 4(b). The reversible multi-level switching of the photonic attenuator was achieved by applying PVAFD, as shown in Fig. 4(c) and Fig. 4(d). The insertion loss and extinction ratio were measured to be 2.91 dB and 39.5 dB, respectively. Furthermore, a multi-level intensity modulation exceeding 180-level (>7 bits) was achieved by applying PVDFA (see Fig. 4(e)). After 1500 switching events, there is no obvious deterioration in the device's performance (see Fig. 4(f)). Further switching events, both with and without an optimized coupling efficiency of the grating coupler after the cyclic measurement, suggest that the observed increase in optical loss can be attributed to a shift in the coupling state.

Although fine-tuning for any target state can always be accomplished via multiple fine-correction pulses akin to the trimming process, the arbitrary state configuration is essential in certain scenarios. The arbitrary state configuration is usually accomplished by applying an amorphization pulse followed by a crystallization pulse. Through this process, we were able to realize a > 3-bit arbitrary state configuration for the GSS4T1-based attenuator. By applying pulses with varying amplitude (varying duration), a total of 7 (13) distinguishable states were achieved (see Fig. 4(g) and (h)).

We have demonstrated a back-end integrated GSS4T1-based photonic attenuator with an ER >39 dB and a multi-level switching >7 bits, thereby forging a pathway for large-scale nonvolatile intensity-modulated PICs.

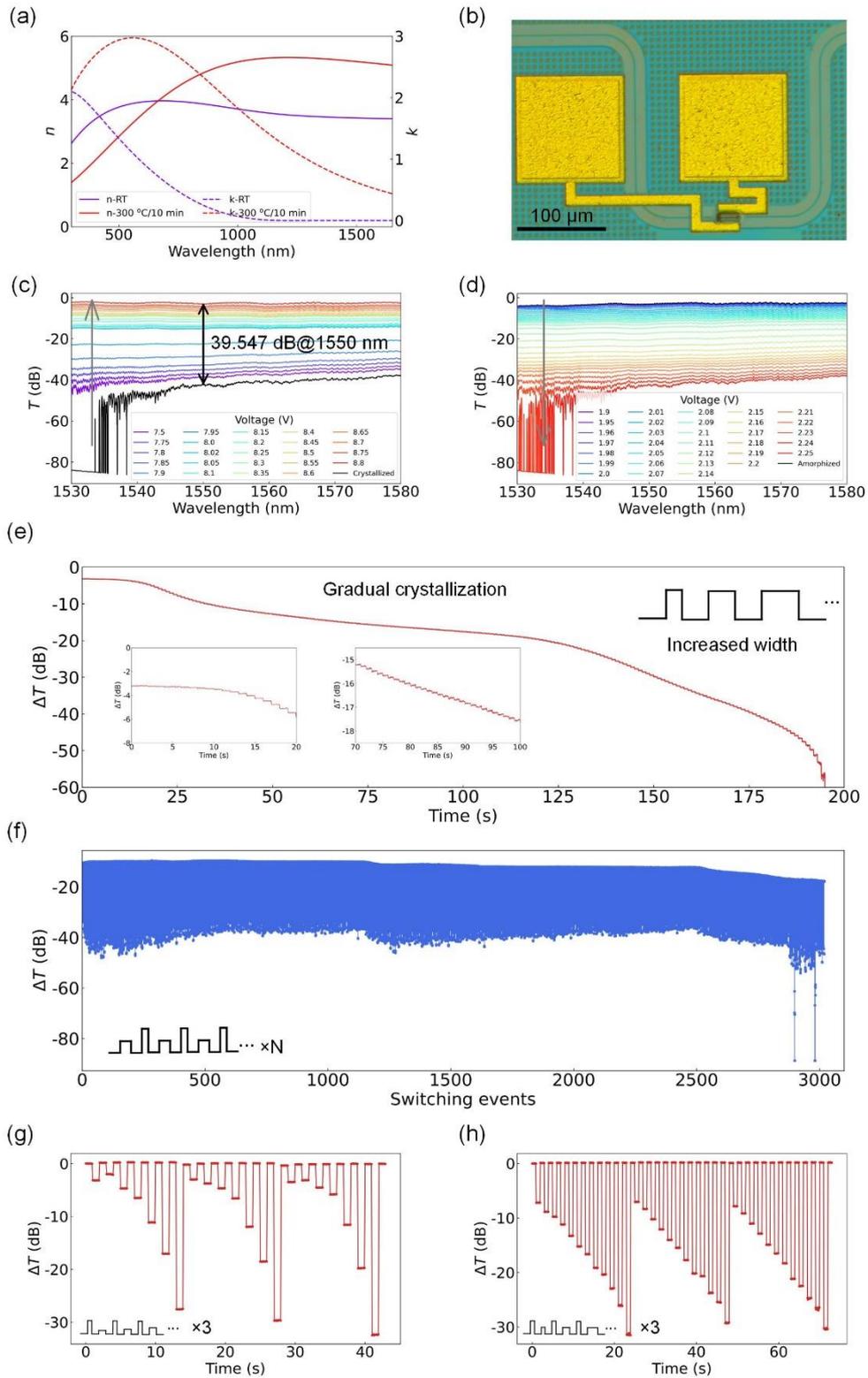

Fig.4 The back-end integration of GSS4T1 with straight waveguide for multi-level intensity modulation. (a) Optical constant of the GSS4T1 film on a silicon substrate before and after annealing. (b)Microscope image of the photonic attenuator. Measured spectra of multi-level amorphization(c) and crystallization(d). (e) Multi-level crystallization induced by applying PVDFA. The insets show gradual crystallization from 0 to 20 and from 70 to 100. (f) 1500 reversible switching events of the photonic attenuator. Arbitrary states configuration induced by varying amplitudes (g) and durations

(h). Transmittance change was all measured at 1550 nm.

**Discussion**

Integrating novel materials into existing passive and active photonic component libraries, enabling the incorporation of unprecedented devices, is crucial for developing next-generation heterogeneous silicon photonic integrated circuits. In this study, we presented a "Zero change" platform for monolithic back-end-of-line integration of phase change materials in silicon photonics. Narrow $SiO_2$ trenches were successfully etched down to the top surface of the Si waveguide core with a SiN etching stop layer on it, enabling the post-deposition of various PCMs for large-scale nonvolatile photonic devices integration without any impact on the foundry-verified photonic devices. The insertion loss of the customized trench is lower than 0.09 dB, allowing large-scale integration in photonic networks.

Two kinds of chalcogenide phase change materials (PCMs), $Sb_2Se_3$ and $Ge_2Sb_2Se_4Te_1$, with completely different nonvolatile modulation capabilities, were monolithically back-end-of-line integrated into silicon photonics. Nonvolatile post-trimming achieved by post-integrated $Sb_2Se_3$ was demonstrated. By electrically fine-tuning the trimming unit to achieve the balance of push and pull voltage, the peak power consumption of a push-pull MZI-type switch was reduced by 48%. The nonvolatile reconfigurable post-fabrication trimming could be applied to various photonic chips against fabrication error. Next, both nonvolatile phase modulation based on $Sb_2Se_3$ and nonvolatile intensity modulation based on $Ge_2Sb_2Se_4Te_1$ employing back-end integration were demonstrated for large-scale nonvolatile programmable photonic networks. The $Sb_2Se_3$-integrated micro-ring switch achieved >5-bit multi-level switching, exhibiting the potential of quasi-continuous switching with electric-pulse-width modulation. Meanwhile, the $Ge_2Sb_2Se_4Te_1$-integrated-waveguide-based broadband attenuator achieved a maximum extinction ratio > 39 dB with >7-bit multi-level modulation. The post-fabricated intensity modulator endured 1500 stable switching cycles without obvious performance degradation.

To our best knowledge, we demonstrate, for the first time, the monolithic back-end-of-line integration of PCMs with chips based on commercial foundry process flow offered by 200-mm commercial silicon photonic foundry and thus the electrically programmable multi-level switching nonvolatile photonic devices. This showcases the feasibility of large-scale-fabricated programmable PCM-based nonvolatile photonic chips, which hold significant potential for low-power, large-scale applications in optical computing, microwave photonics, and optical communication networks. And the "Zero change" monolithically back-end-of-line integration platform could also pave the way to integrate other excellent optoelectronic materials into the future silicon photonic process design kits.

## Methods
### Device characterization
The broadband tunable laser (Santec TSL-550) emits a signal light that is directed to the polarization controller (PC) for adjusting polarization, and subsequently couples with the device under test (DUT) through the double-end grating coupler. The output optical power of the DUT is measured by an optical power meter ( MPM-210 ). The electric pulse is generated by an arbitrary waveform generator ( SDG7052A ) and applied to the electrode of the DUT through a radio frequency probe.

### Data Availability
All the data supporting this study are available in the paper and Supplementary Information. Additional data related to this paper are available from the corresponding authors upon request.

### Acknowledgments
This work was supported by the National Natural Science Foundation of China (91950204，62105287，61975179，92150302), the National Key Research, Development Program of China (2021YFB2801300，2019YFB2203002), and the Zhejiang Provincial Natural Science Foundation of China (LD22F040002). The authors would like to acknowledge the fabrication support from the Institute of Microelectronics of the Chinese Academy of Sciences, ZJU Micro-Nano Fabrication Center at Zhejiang University, and Westlake Center for Micro/Nano Fabrication at Westlake University. The authors would also like to thank Liming Shan for his help in chalcogenide thin-film depositions and Mengxue Qi for her help in device fabrication.

**Author contributions**

H.L. conceived the idea. M.W., K.X., and B.T. carried out fabrication, measurement setup construction, and device testing. J.L. contributed to the fabrication. Y.Y. assisted in device testing. P. Z. performed passive device characterization. Y.W. performed the protective layer deposition. K.B. assisted in the wafer dicing. K.L. developed the PCMs films. Z.C. developed and deposited the PCMs film. H.M. assisted with measurement setup construction. C.S. contributed to the passive devices design. R.L. contributed to the tape-out of GDS files. H.L., L.L., J.L., M.L. supervised the research. All authors contributed to technical discussions and writing the paper.

**Competing interests**

The authors declare no competing financial interests.